# Passive frequency conversion of ultraviolet images into the visible using perovskite nanocrystals

Jad Salman[1], Mahesh K. Gangishetty[2,6], Bryan E. Rubio-Perez[1], Demeng Feng[1], Zhaoning Yu[1,3], Zongzhen Yang[4], Chenghao Wan[1,5], Michel Frising[1], Alireza Shahsafi[1], Daniel N. Congreve[6], Mikhail A. Kats[1,3,5]

[1]Department of Electrical and Computer Engineering, University of Wisconsin–Madison, Madison WI 53706, [2]Department of Chemistry, Mississippi State University, Starkville, MS 39762, [3]Department of Physics, [4]Department of Computer Sciences, [5]Department of Materials Science and Engineering, University of Wisconsin–Madison, Madison WI 53706, [6]Rowland Institute at Harvard University, Cambridge, MA 02142

**Abstract:**

We demonstrate a passive down-conversion imaging system that converts broadband ultraviolet light to narrow-band green light while preserving the directionality of rays, and thus enabling direct down-conversion imaging. At the same time our system has high transparency in the visible, enabling superimposed visible and ultraviolet imaging. The frequency conversion is performed by a subwavelength-thickness transparent downconverter based on highly efficient $CsPbBr_3$ nanocrystals incorporated into the focal plane of a simple telescope or relay-lens geometry. The resulting imaging performance of this down-conversion system approaches the diffraction limit. This demonstration sets the stage for the incorporation of other high-efficiency perovskite nanocrystal materials to enable passive multi-frequency conversion imaging systems.

**Introduction**

The typical human eye is sensitive to wavelengths between roughly 400 and 700 nm [1,2]. However, beyond the visible—*e.g.*, in the infrared and ultraviolet—many objects convey unique signatures to the reflected, transmitted, and/or emitted spectra. For example, ultraviolet imaging has applications for forensic imaging [3], photography for remote sensing of animals and plants [4,5], and finding electrical faults in high-power systems through detections of corona discharge [6]. Taking inspiration from nature, imaging using ultraviolet light can also be used to detect camouflaged objects. For example, caribou vision has evolved to be sensitive to ultraviolet wavelengths, allowing for increased contrast in imaging what would otherwise be well-camouflaged predators [7]. In the broadest sense, the ability to visualize scenes comprised of spectral information beyond the response of the human eye requires a form of frequency conversion to map invisible light to the visible while preserving the directionality of rays to enable imaging.

Commercial systems—such as ultraviolet, infrared, and hyperspectral cameras—used for imaging outside the visible range are active power-consuming electro-optical instruments that typically decouple the ability to directly view scenes, relying on separate displays.

Unlike the active devices noted above, passive frequency-conversion imaging systems utilize no external electrical sources or detectors. Commercialized passive frequency-conversion devices are predominately used for profiling lasers (*e.g.,* the Newport™ LBP2-UVIMG ultraviolet image converter) and rely on phosphor materials as the frequency-converting medium. Phosphors in powder form are not transparent in the visible [8,9], and such systems utilize the phosphor as a screen upon which to project the ultraviolet images or deposit the phosphor directly onto the imaging array pixels [10], eliminating the ability to simultaneously transmit visible images. Recent work on developing thin films of transparent crystalline phosphors [11] or phosphors embedded within transparent glasses [12] could be promising for applications in frequency-conversion imaging. Similarly, down-converting nanocrystals, or quantum dots, of organic and inorganic semiconducting materials are a promising avenue for passive frequency-conversion imaging due to their simple synthesis, tunability of emission, and high quantum efficiency [13–18]. In particular, perovskite nanocrystals are a new down-converting material of interest with superior optical properties such as high color purity and color tunability compared to organic phosphors [19].

In this paper, we demonstrate a passive frequency-conversion imaging system that converts broadband ultraviolet light to narrowband green light and is mostly transparent across the visible range—enabling simultaneous visible and ultraviolet imaging with the naked eye or a conventional visible camera. Our conversion system utilizes a subwavelength-thick transparent downconverter film made of highly efficient cesium lead bromide ($CsPbBr_3$) perovskite nanocrystals incorporated into the focal plane of a simple telescope or relay lens geometry. This proof of concept sets the stage for incorporating other high-efficiency downconverter materials to enable passive multispectral frequency conversion to enhance human spectral perception or to enable inexpensive multispectral imaging.

**Passive down-conversion imaging**

A conceptual example of our passive down-conversion imaging system is described in figure 1. An arctic wolf has a fur coat that appears white in daylight due to the broadband scattering of visible light—useful for camouflage in the wolf's snowy habitat. However, the wolf's coat is absorbing in the ultraviolet, in contrast to the snow which remains scattering into the ultraviolet [7]. Thus, the ability to see in the ultraviolet provides an evolutionary advantage to prey animals like caribou. Inspired by this natural adaptation, we sought to create a passive apparatus (i.e., one which requires no external power source or digital logic) to enable, e.g., simultaneous ultraviolet and visible vision for humans (figure 1(d-e)).

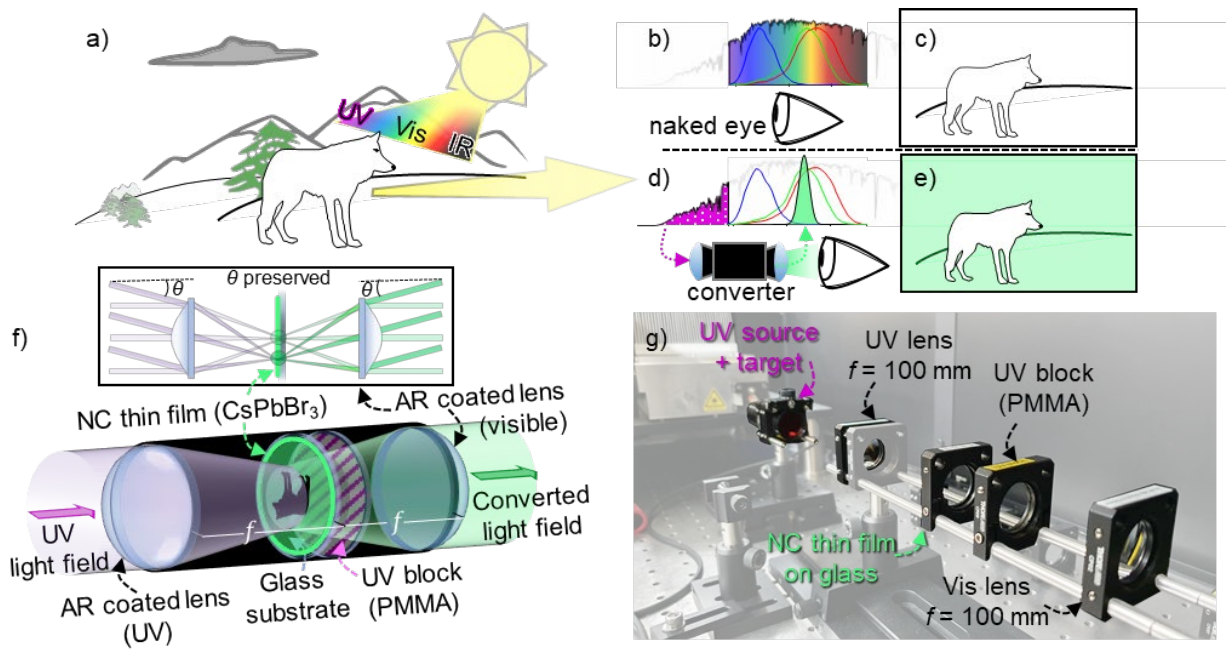

**Figure 1: Passive down-conversion imaging**. a) An arctic wolf's fur coat scatters and reflects sunlight, which is spectrally broad. However, ultraviolet (UV) frequencies are strongly absorbed by its fur compared to the surrounding snow. To an unaided human eye (b,c), the wolf blends in with the snowy background since human vision is only sensitive to wavelengths within the narrow visible band, 400 – 700 nm (highlighted rainbow region of solar spectrum in the inset), and is limited to three spectral bins with broad overlapping sensitivities (red, green, and blue curves inset b). d) Our device converts ultraviolet light into the visible within a narrow wavelength band (here, in the green) while maintaining the directionality of rays, thus enabling passive frequency-conversion imaging. f) A schematic of our converter system, which uses a thin-film coating of $CsPbBr_3$ nanocrystals (NCs) as the frequency-conversion medium, with the film facing the objective lens. The focal arrangement images far-away scenes with 1:1 magnification. e) An image of our experimental setup with all relevant optical elements labeled. A collimated UV diode

($\lambda_o$ = 340 nm, FWHM = 11 nm) is used as our illumination source, with the frequency-converted light centered at 510 nm.

We demonstrated a passive down-conversion imaging system by incorporating a visibly transparent thin-film perovskite nanocrystal material ($CsPbBr_3$) into the intermediate focal plane of a refracting telescope (figure 1(f,g)). Since the converter material cannot by itself maintain angular directionality of incident rays of light, *i.e.*, the converted light at each point on the $CsPbBr_3$ film luminesces incoherently in all directions, the objective lens was necessary to form an intermediate image onto the converter, which down-converts the ultraviolet components of the image (figure 1(f) inset). The eyepiece lens, placed one focal length away from the converter, remaps the intermediate image back to propagating rays with the same angles of incidence as the incident ultraviolet rays, allowing for the converted ultraviolet images and the directly transmitted visible images to be viewed simultaneously. We note that, in principle, the introduction of down-converted light may lead to difficulty distinguishing between converted and visible images (i.e., visible/ultraviolet metamers). This can be avoided for binocular vision if the down-conversion imaging system is applied only to one eye [20].

**Results and discussion**

$CsPbBr_3$ nanocrystals have near-unity photoluminescence (PL) quantum yield when excited with near-ultraviolet light [21] and emit in a narrow wavelength range, making this material platform an ideal candidate for down-conversion imaging. We synthesized undoped $CsPbBr_3$ nanocrystals, with individual crystal sizes of ~10 nm, via the process described in ref. [22] under Experimental Procedures for undoped perovskite nanocrystals, and in ref. [23]. The nanocrystals were suspended in a solution of hexane at a concentration of ~30 mg of nanocrystals per 1 ml of solvent. Based on prior results from nanocrystals synthesized using the synthesis process in ref. [22], the PL spectrum was expected to be centered at a wavelength of 510 nm with a full width at half maximum (FWHM) of 20 nm. Under typical white-light illumination, the solution has a green-yellow tint (figure 2(a)). When 340-nm ultraviolet light (~50 mW) is incident onto the nanocrystal solution, a vibrant green emission is clearly visible (figure 2c) and dominates even under mixture of normal white-light conditions and ultraviolet illumination (figure 2(b)). The solution was spin cast onto two 25-mm-diameter BK7 float glass substrates at 2000 RPM for 30 seconds to create an ~80-nm-thick nanocrystal layer.

Perovskite nanocrystals are known to be unstable when exposed to moisture, heat, and prolonged ultraviolet illumination. Several works have showed improved stability and performance longevity when a polymer layer seals the nanocrystal film or is used as a matrix for the nanocrystals [16,22,24–29]. Zhu et al. [27] reported stable performance of $CsPbBr_3$ in a PMMA matrix after 12 hours immersed in water, while Li et al. [28] showed good stability after months of immersion in room-temperature water as well as after hours of near-boiling temperatures for perovskite nanocrystals grafted with PMMA ligands. And PMMA nanofibers with embedded $CsPbBr_3$ nanocrystals have been shown to have very stable photoluminescence under hours of continuous direct ultraviolet illumination [29]. Our device is expected to operate well within these extremes. To test the impact of incorporating a polymer protective capping layer on imaging and efficiency, we subsequently coated one sample with an ~70-nm-thick layer of PMMA (4% w/V PMMA in anisole).

The 80-nm-thick nanocrystal layer is far below the depth of field of the imaging optics (here, 10.7 μm), and therefore should enable down-conversion imaging at the diffraction limit, as described below. The PL spectrum of the coated samples was measured, with the peak emission at ~510 nm for the nanocrystal-only sample, while the PMMA-capped sample had its emission slightly shifted to ~520 nm (see supplemental). Both samples remained highly transparent in the visible (figure 2(d)).

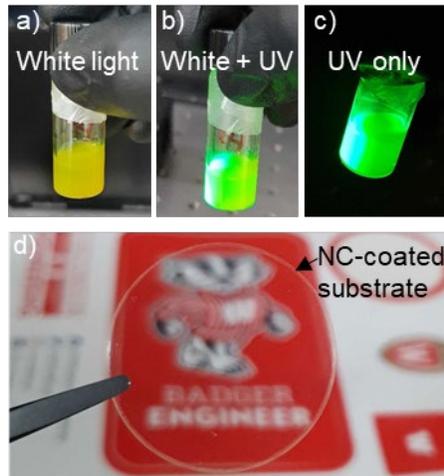

**Figure 2: CsPbBr₃ downconverter.** a) A vial of CsPbBr$_3$ nanocrystals suspended in hexane under white-light illumination, (b) exposed to both white light and ~50 mW of UV light centered at λ = 340 nm, (c) and only under UV illumination. The vibrant green emission of converted light is centered at λ = 510 nm. (d) The 80-nm-thick nanocrystal film deposited on BK7 glass is highly transparent in the visible (the blurry image is because the photo is focused on the sample rather than the background).

In our down-conversion imaging system, the thin-film samples were placed in the intermediate image plane between two 100-mm-focal-length spherical lenses. The objective lens was anti-reflection coated to maximize transmission of ultraviolet light (Thorlabs LA4380-UV, NA = 0.13). The eyepiece lens was anti-reflection coated to maximize transmission in the visible (Thorlabs LA1509-A, NA = 0.13). An ultraviolet blocking filter (1.5-mm-thick polycarbonate) was placed after the converter to prevent transmission of ultraviolet light to the eyepiece (figure 1(g)).

We characterized the imaging resolution of our conversion system by imaging a standard 5 mm × 5 mm Air Force resolution target (Thorlabs R1DS1N) (figure 3(a)) illuminated from the backside with a 50-mW narrow-band ultraviolet light-emitting diode (LED, Thorlabs M340L4), with the output centered at 340 nm and a FWHM of 11 nm. The source was collimated using an antireflection-coated fused-silica aspheric lens (Edmund Optics, 33-953, NA = 0.69). A visible-blocking ultraviolet bandpass filter (Asahi Spectra XRR0340) was placed after the collimator lens to eliminate unwanted visible light from the LED. The divergence of the collimated beam entering the objective lens was approximately 3.3°. The resolution target was placed $2f$ (200 mm) away from the objective, such that the image magnification on the nanocrystal plane was 1:1. The nanocrystal films were oriented facing the objective side. The eye piece lens was placed $1f$ (100 mm) distance away from the nanocrystal plane such that the converted image was collimated at the output. A digital camera (Nikon D5600) focused to infinity was placed at the end of the eyepiece lens.

Figure 3(b) shows an image of the down-converted resolution target as captured through our setup under a normal white-light ambient background. A separate color logo was placed in the same plane as the resolution target demonstrating how our down-conversion imaging system can simultaneously convert ultraviolet images into the visible while preserving the ability to directly image visible scenery.

We then imaged the nanocrystal-only and PMMA-capped samples under ultraviolet illumination, without any ambient lighting (figure 3(d,e)), and compared the images to the resolution target illuminated by a 520-nm-wavelength reference diode (figure 3(c)) to eliminate resolution discrepancies due to chromatic aberrations. The depth of focus (DOF) of the objective lens—approximated as $\lambda/(2NA^2)$ in air, where $\lambda$ is the wavelength of light focused by the objective—defines the tolerance along the focal length of the lens where the image is sharp and maintains its resolution (see supplemental). Thus, a down-converting film that is thicker than the DOF of the imaging system can result in degradation of imaging, because emission from the film outside of the DOF will result in blurring of the final converted image. Our objective lens (NA = 0.13) provided a DOF of 10.7 μm at a wavelength of 340 nm. With our nanocrystal film thicknesses being several orders of magnitude thinner, any aberrations in imaging quality were expected to be exclusive of DOF effects and predominately from the quality of the converter film itself.

As seen in figure 3(c-e) insets, the smallest resolvable feature—defined here as a least a 50% modulation in signal intensity between horizontal lines and spaces (white curves)—was nearly the same between the reference diode and the nanocrystal-only film (pink dashed boxes, figure 3(c,d) insets). The pitch of this group-4-element-6 horizontal line pair was 35 μm. For the PMMA-capped sample, the smallest resolvable feature (blue dashed boxes, figure 3(e) inset) was the group-4-element-4 line pair, with a pitch of 44 μm. The slight degradation in resolution for the PMMA-capped sample was attributed to the nonuniformity of the capping layer and nanocrystal layer; this nonuniformity can be seen in the image as radial streaking of bright and dim regions.

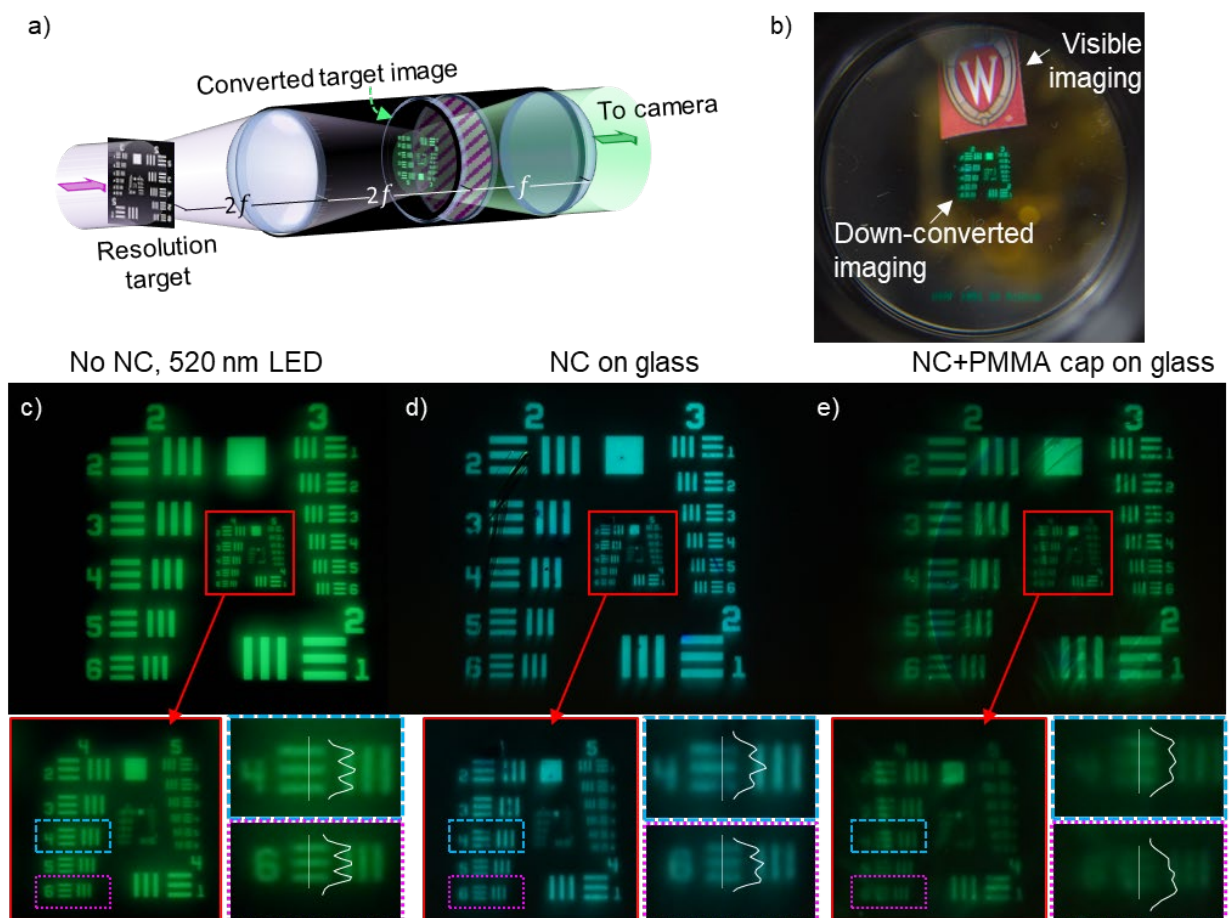

**Figure 3: Down-conversion imaging.** (a) A schematic of our setup used to image a US Air Force resolution target. The 50-mW ultraviolet source illuminated the backside of the target, which was defined by openings in a chrome mask. The target was placed 2*f* away from the objective lens to form a 1:1 (unmagnified) image at the nanocrystal plane. (b) A photo through our down-conversion imaging system showing the converted image of the resolution target. The surrounding background is visible due to ambient white light in the scene. A color university logo is placed in the same plane as the resolution target to show our system's ability to simultaneously

convert ultraviolet images into the visible and preserve the ability to image the surrounding visible scenery. (c) A baseline image of the resolution target illuminated from behind with a green LED and no nanocrystal (NC) converter in place. Pairs of horizontal and vertical lines and spaces are denoted by group number and element number, *e.g.*, the far lower-right corner set of lines and spaces is labeled as group 2 element 1. (d) Frequency-converted images of the resolution target with the nanocrystal film on glass, and (e) PMMA-capped nanocrystal film on glass samples in place. A shift in the color was noted for the PMMA-capped sample and had a peak PL emission at 520 nm. (c-e insets) the minimum resolvable features for both the baseline image and the nanocrystal-only film are the group-4-element-6 lines (pink inset region) with the horizontal bars being distinguishable (white line = signal intensity profile). The minimum resolvable features of the PMMA-capped nanocrystal film are the group-4-element-6 lines (blue inset region). Only minimal degradation to imaging resolution is noted compared to the baseline performance.

Next, we determined the system efficiency of our setup, $\eta$, defined as the number of photons emerging from the eyepiece divided by the ultraviolet photons measured at the converter plane (figure 4(a)). The lenses were repositioned to be exactly one focal length apart from the converter plane to focus incoming collimated light to a point on the surface of the nanocrystal film. As shown in figure 4(a), an extra lens (Thorlabs LA1131, NA = 0.25) was added just after the eyepiece lens (denoted as "eye") to focus the converted light onto a silicon photodiode power meter (Thorlabs S130VC) to measure the amount of light that is converted by the system and can enter the eye. This value was then normalized by the power of ultraviolet light that enters the system and is focused onto the nanocrystal plane. We scaled the value of the ultraviolet power by 2/3 ((340 nm) / (510 nm)) to account for differences in photon energies so that $\eta$ represents the number of converted photons, *i.e.*, $\eta$ = 100% means all incident photons are converted and captured by the optics.

The two major factors that determine $\eta$ are the optics used to collect converted light and the performance of the nanocrystal samples, *i.e.*, the amount of ultraviolet light that is absorbed, the PL quantum yield, and the amount of converted light that can escape, or outcouple, from the material.

First, the absorptance of 340-nm ultraviolet light by the nanocrystal film was defined as $\alpha = 1 - T_{film}$. $T_{film}$ is the nanocrystal-film transmittance and was calculated by dividing the measured transmitted ultraviolet power through the entire system with the nanocrystal samples in place by the ultraviolet power through the system with no nanocrystal layer (*i.e.*, just glass or glass and the 70 nm PMMA capping layer only). Scattering by the films were assumed to be minimal, but it is noted that scattering can inflate the absorptance values for particularly hazy films. The absorptance was measured to be 7.5% $\pm$ 0.8% for the nanocrystal-only sample and 6.2% $\pm$ 0.6% for the PMMA-capped sample (see Supplementary Information). The uncertainties were calculated from the power-meter uncertainty specifications.

As is known for many high-index light-emitting materials, the amount of light that can outcouple into free space is often significantly less than the internal quantum yield—near unity for these materials—due to effects such as total internal reflection at the boundary between the high-refractive-index active layer film and surrounding low-index materials, and reabsorption of photons by the perovskite material [30–33]. Thus, the outcoupling efficiency is necessarily less than unity for an unstructured thin film.

We determined that roughly 6% of the emitted light could outcouple into free space towards the eyepiece lens by performing 3D full-wave simulations of a point source—representing a localized nanocrystal—emitting at 510 nm centered within an 80-nm-thick film of $CsPbBr_3$ (n = 2.1) [34] on top of BK7 glass substrate (n = 1.54) (see Supplementary Information). Using the simulated outcoupling efficiency and the measured absorption, we calculated that 0.4% of the ultraviolet photons incident on the film are ultimately converted into visible photons that are reemitted into free space. After accounting for transmittances and numerical apertures for all optical components in the system (see Supplementary Information) the *expected* system efficiency ($\eta_{expected}$) is approximately 0.001%. We measured the $\eta$ with the nanocrystal-only sample to be 0.0013% $\pm$ 0.0001%, and with the PMMA-capped nanocrystal sample to be 0.00086% $\pm$ 0.00007%, both in good agreement with $\eta_{expected}$.

In figure 4(b), we plotted the projected improvements to the system efficiency, η, with various potential enhancements to the down-conversion imaging system compounded together. First, by eliminating all reflection losses from optical elements with antireflection coatings, η can be nearly doubled. Incorporating a back reflector that passes ultraviolet and visible but redirects backwards-propagating converted emission in the forward direction (e.g., a reflecting dielectric notch filter) just before (or physically deposited on top of) the nanocrystal films can also improve system efficiency by 50%, assuming an initially isotropic emission profile from the film. Further improvements to the nanocrystal film performance, such as increasing absorptance and outcoupling efficiency, will also yield orders-of-magnitude efficiency gains. This can be readily implemented by thickening the down-converter film either by embedding concentrated nanocrystals in a polymer matrix or by depositing multiple alternating layers of pure nanocrystals/polymeric matrices using a mixture of solvents (*e.g.*, a polymeric matrix dissolved in toluene for one layer, followed by nanocrystals dispersed in hexane for the other). Although current highest-reported outcoupling efficiencies have reached 20% [30,32], it is possible to exceed this with structuring of the film or substrate surface to reduce waveguiding effects from interface reflections [33,34]. Finally, the use of higher-NA collection optics (e.g., increasing from NA = 0.13 to NA = 0.8) is readily achievable and will enable much more of the emission to be captured, significantly increasing η. As shown in figure 4(b), the cumulative effect of these systemic improvements can increase η by many orders of magnitude.

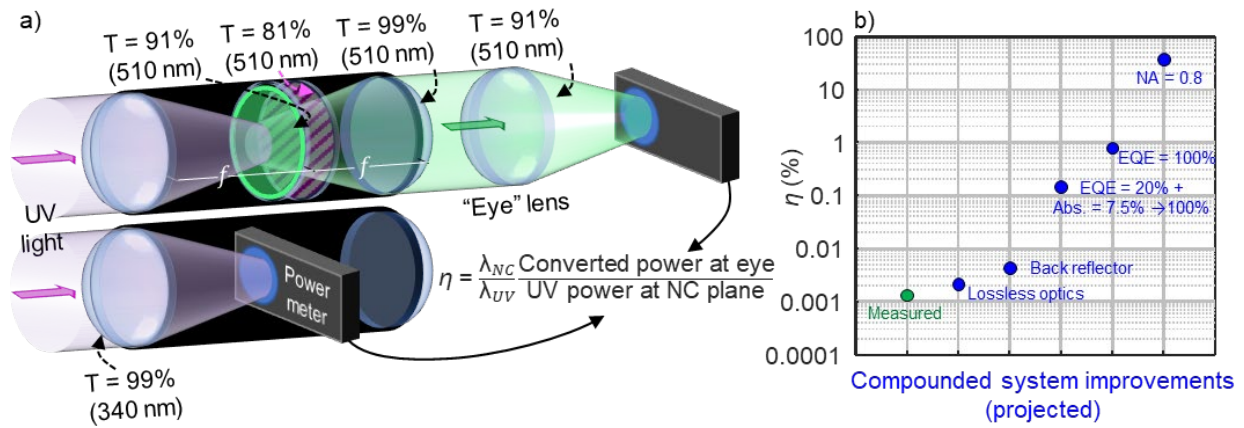

**Figure 4: Converter system efficiency.** a) The system efficiency, *η*, is experimentally determined by measuring the power of the converted light at the eye or visible camera, normalized by the ultraviolet power that is focused onto the nanocrystal plane. The transmittance of each optical component is noted for the respective wavelengths. b) The measured *η* (green dot), and calculated *η* assuming a variety of improvements to the down-conversion system, compounded at each point. For instance, the use of lossless optics would double *η* from the measured value. Furthermore, incorporating a back reflector between the nanocrystal structure and objective lens that can redirect backwards-propagating emission—which is typically lost—towards the eyepiece, further doubles of efficiency, and so on. Orders-of-magnitude improvements in system efficiency are possible with increasing external quantum efficiency, enhanced absorption, and the incorporation of larger-NA collecting optics.

**Conclusion**

We demonstrated a passive down-conversion imaging system based on thin films of efficient $CsPbBr_3$ nanocrystals to enable ultraviolet imaging using a visible camera or the naked eye. The nanocrystal films are thin and highly transparent, allowing for simultaneous diffraction-limited imaging of visible and down-converted images from the ultraviolet wavelengths. We quantified the system efficiency of our devices given two nanocrystal samples—with and without a PMMA cap to protect the nanocrystal film—and extrapolated expected efficiencies given reasonable improvements in the material and optical-system performance. Alternative nanocrystal materials can be incorporated into the system to allow multispectral down-conversion [22,38,39] and/or up-conversion imaging [40], thus expanding human spectral perception aided by only passive optics into both the ultraviolet and the infrared.

## Acknowledgements

MAK acknowledges support from the Air Force Office of Scientific Research, award FA9550-18-1-0146. DNC and MKG acknowledge the support of the Rowland Fellowship at the Rowland Institute at Harvard.

34. W. Yan, W. Yan, L. Mao, P. Zhao, A. Mertens, A. Mertens, S. Dottermusch, H. Hu, Z. Jin, Z. Jin, B. S. Richards, and B. S. Richards, "Determination of complex optical constants and photovoltaic device design of all-inorganic $CsPbBr_3$ perovskite thin films," Opt. Express, OE **28**, 15706–15717 (2020).
35. K. Lin, J. Xing, L. N. Quan, F. P. G. de Arquer, X. Gong, J. Lu, L. Xie, W. Zhao, D. Zhang, C. Yan, W. Li, X. Liu, Y. Lu, J. Kirman, E. H. Sargent, Q. Xiong, and Z. Wei, "Perovskite light-emitting diodes with external quantum efficiency exceeding 20 per cent," Nature **562**, 245–248 (2018).
36. J. M. Richter, M. Abdi-Jalebi, A. Sadhanala, M. Tabachnyk, J. P. H. Rivett, L. M. Pazos-Outón, K. C. Gödel, M. Price, F. Deschler, and R. H. Friend, "Enhancing photoluminescence yields in lead halide perovskites by photon recycling and light out-coupling," Nature Communications **7**, 13941 (2016).
37. G. Mei, D. Wu, S. Ding, W. C. H. Choy, K. Wang, and X. W. Sun, "Optical Tunneling to Improve Light Extraction in Quantum Dot and Perovskite Light-Emitting Diodes," IEEE Photonics Journal **12**, 1–14 (2020).
38. Z. Gan, H. Xu, and Y. Hao, "Mechanism for excitation-dependent photoluminescence from graphene quantum dots and other graphene oxide derivates: consensus, debates and challenges," Nanoscale **8**, 7794–7807 (2016).
39. X.-B. Shen, B. Song, B. Fang, A.-R. Jiang, S.-J. Ji, and Y. He, "Excitation-wavelength-dependent photoluminescence of silicon nanoparticles enabled by adjustment of surface ligands," Chemical Communications **54**, 4947–4950 (2018).
40. T. N. Singh-Rachford and F. N. Castellano, "Photon upconversion based on sensitized triplet–triplet annihilation," Coordination Chemistry Reviews **254**, 2560–2573 (2010).

Supplementary information:

**Passive frequency conversion of ultraviolet images into the visible using perovskite nanocrystals**

Jad Salman[1], Mahesh K. Gangishetty[2,6], Bryan E. Rubio-Perez[1], Demeng Feng[1], Zhaoning Yu[1,3], Zongzhen Yang[4], Chenghao Wan[1,5], Michel Frising[1], Alireza Shahsafi[1], Daniel N. Congreve[6], Mikhail A. Kats[1,3,5]

[1]Department of Electrical and Computer Engineering, University of Wisconsin–Madison, Madison WI 53706, [2]Department of Chemistry, Mississippi State University, Starkville, MS 39762, [3]Department of Physics, [4]Department of Computer Sciences, [5]Department of Materials Science and Engineering, University of Wisconsin–Madison, Madison WI 53706, [6]Rowland Institute at Harvard University, Cambridge, MA 02142

## 1. Spectra of CsPbBr$_3$ thin films photoluminescence and reference LED diode

We measured the photoluminescence (PL) spectra of the two CsPbBr$_3$ thin-film samples used for down-conversion imaging (figure S1). The samples were illuminated with a 50-mW ultraviolet source with a peak wavelength at 340 nm. The samples were measured using an Ocean Optics (OO) FLAME-S-VIS-NIR-ES grating spectrometer coupled to an optical fiber. Located at the end of the optical fiber was an OO CC-3 Cosine Corrector (opaline glass diffuser) to collect signal from a 180° field of view. The instrument was calibrated to a known light source (OO, HL-3P-CAL, 350-1100 nm). The fiber probe was placed a few millimeters away from the illuminated spot (figure S1). A slight ~10-nm red shift in the PL spectrum was noted between the nanocrystal film that was capped with ~70-nm PMMA and the uncapped nanocrystal-film. Furthermore, we measured the emission from a reference green LED use to baseline the imaging resolution of the optical setup (figure S1). The peak wavelength was 520 nm with a full width at half maximum of 30 nm, similar to the peak wavelength of our thin-film PL.

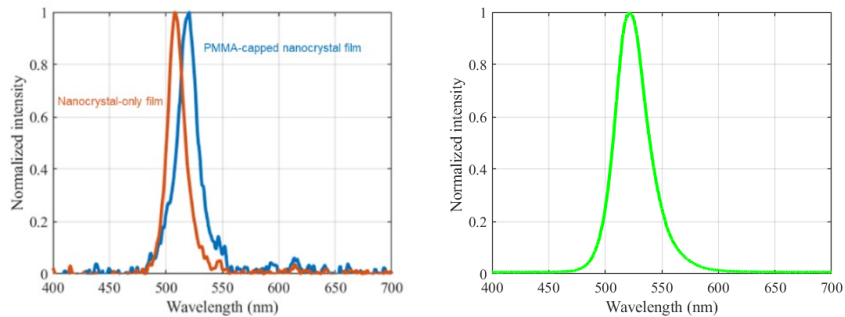

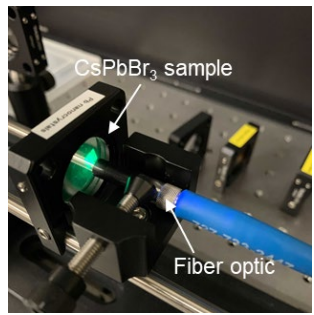

**Figure S1:** (left) Photoluminescence measurements of the two thin films characterized in the main text. The nanocrystal-only sample has a peak emission wavelength of 510 nm and the PMMA-capped nanocrystal sample has a peak emission wavelength of 520 nm. The full width at half maximum (FWHM) is 20 nm for both samples. (right) The measured emission spectrum from the reference green LED used in the imaging test in figure 3c to baseline the imaging performance of the optics in our system. The spectrum has a peak wavelength at 520 nm and a FWHM of 30 nm. (bottom) An image of the photoluminescence measurement setup with the spectrometer optical fiber placed near a thin-film sample.

## 2. Depth of Focus calculation

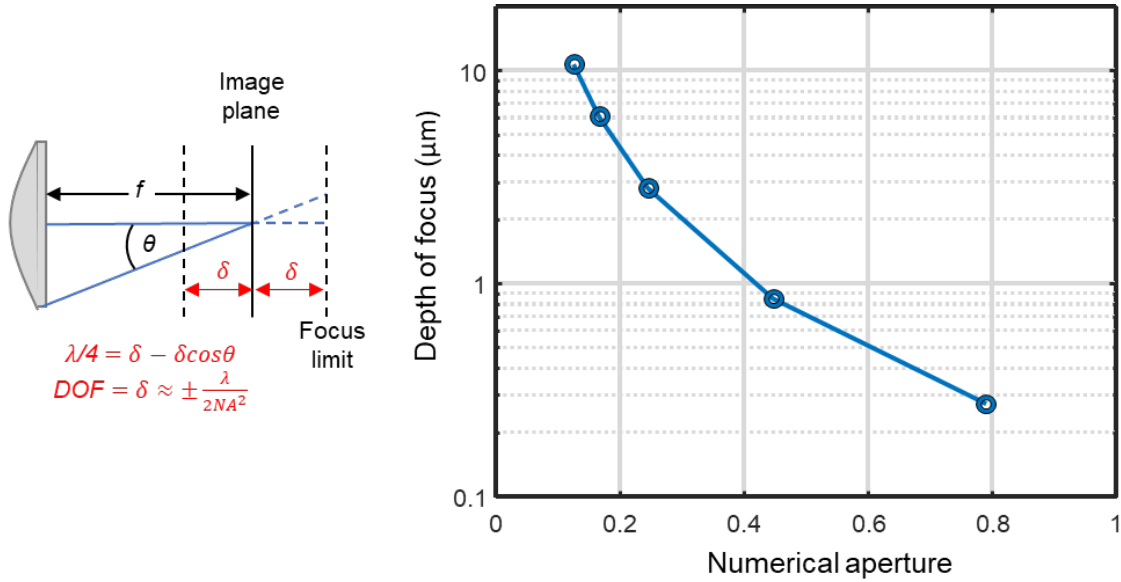

**Figure S2: Depth of focus versus numerical aperture.** Calculations of depth of focus of the objective lens versus numerical aperture. Numerical aperture is defined as $NA = n\sin(\theta)$, where $\theta$ is the maximum angle defined by the focal length and diameter of a spherical lens and $n$ is the refractive index surrounding the lens—in this case air. Depth of focus, defined as DOF = $\lambda/(2NA^2)$, where $\lambda$ is the wavelength of the focused light, describes the region along the optical axis where the path-length difference between focused rays does not exceed $\lambda/4$ in the paraxial approximation as described in the left part of the figure. This region is where a focused image remains sharp. The objective lens used in our experiment had a focal length of 100 mm and an NA of ~0.13, giving a DOF of 10.7 µm. Our down-converting films were deposited with a thickness of 80 nm, much smaller than the DOF of our system. Even for high-NA optics with NA = 0.8, the DOF = 180 nm, and thus we can expect to maintain imaging performance irrespective of defocusing effects with our existing down-converting layer.

## 3. Absorptance of the nanocrystal films

As described in the main paper, the absorptance, $\alpha$, at a wavelength of 340 nm for the nanocrystal-only and PMMA-capped nanocrystal samples was determined experimentally using the optical setup in figure S2. $\alpha$ was calculated as $\alpha = 1 - T_{film}$, where $T_{film}$ is the transmittance of ultraviolet light through the nanocrystal film. The transmittance was found by measuring the power of ultraviolet light through the entire system with the nanocrystal film in place divided by the power of ultraviolet light without the nanocrystal film (*i.e.*, with just a bare BK7 glass substrate or just the BK7 glass substrate with the ~70-nm PMMA capping layer, see figure S3). An ultraviolet bandpass filter was placed just before the power meter to remove any converted light. The absorptance was measured to be 7.5% ± 0.8% for the nanocrystal-only sample and 6.2% ± 0.6% for the PMMA-capped sample. The uncertainties were calculated based on the instrument specifications of the power meter. Reflectance from the ~80-nm nanocrystal film was found to be negligible, since the thickness is nearly $\lambda/(2n)$ where n ≈ 2.1 at $\lambda$ = 340 nm [1], allowing for destructive interference of the reflected light for normally incident light.

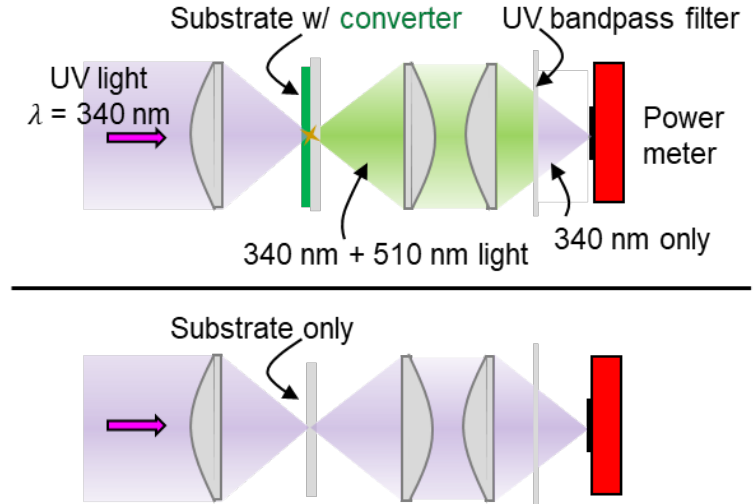

**Figure S3: Ultraviolet absorption measurement.** Absorptance, $\alpha$, is defined here as 1-$T_{film}$ where $T_{film}$ is the transmittance of ultraviolet light at 340 nm through the CsPbBr$_3$ nanocrystal film. (top) A diagram of the measurement of ultraviolet power through the down-conversion imaging setup with the nanocrystal film sample in place. An ultraviolet bandpass filter (Asahi Spectra Inc., XHS0350 ultraviolet pass filter) is placed in front of the power meter to block all converted light. (bottom) The measurement is repeated with the same setup as before, but with only a bare substrate without the nanocrystal film in place of the original sample. $T_{film}$ is the ratio of the two power measurements. The same measurement is repeated with the PMMA-capped nanocrystal sample.

### 4. Calculation of expected system efficiency

The expected system efficiency, $\eta_{expected}$, is defined as $P_{captured}/P_{ultraviolet} \times \lambda_{green}/\lambda_{ultraviolet}$, where $P_{captured}$ is the calculated power of converted (*i.e.*, visible) light captured and propagated through the optical system, $P_{ultraviolet}$ is the measured power of ultraviolet light incident on the CsPbBr$_3$ nanocrystal film, and $\lambda$ is either the wavelength of the green (510 nm) or ultraviolet (340 nm) light. We reasonably assume that light emitted by the nanocrystal film can be approximated as isotropic (figure S4). Therefore, $P_{captured}$ is calculated as:

$$P_{captured} = \underbrace{\frac{(P_{ultraviolet})(QY_{out})(\alpha)}{4\pi}}_{\text{Radiant intensity (W/sr)}} \underbrace{(2\pi(1-\cos(\theta_{NA}))}_{\text{Conical area captured (sr)}} \underbrace{(T_{glass})(T_{block})(T_{eyepiece})(T_{eye})}_{\text{Transmittances}},$$

Where $QY_{out}$ is the outcoupling efficiency, $\alpha$ is the ultraviolet absorptance of the nanocrystal film, $\theta_{NA}$ is the half angle of the conical area captured by the eyepiece lens as defined by the numerical aperture, and $T_{glass}$, $T_{block}$, $T_{eyepiece}$, and $T_{eye}$ are the transmittance at 510-nm light for each optical element in the path, as defined in figure S4. The $\eta_{expected}$ shown in figure 4b are listed in table S1.

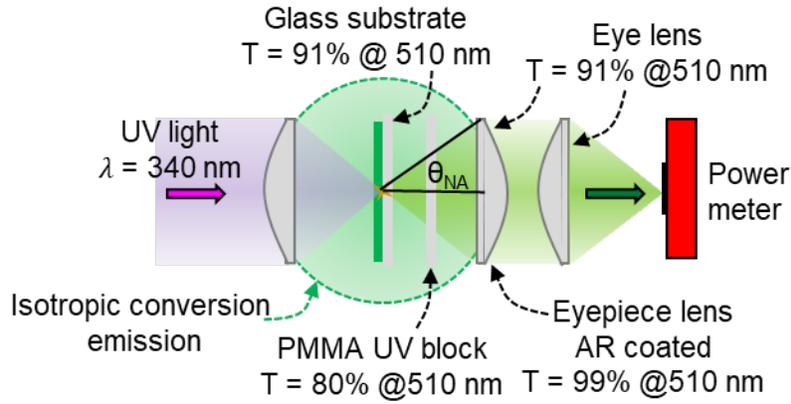

**Figure S4: Captured converted power.** A diagram showing the power-measurement setup for capturing the down-converted light from the nanocrystal film. The calculation of $P_{captured}$ is made assuming isotropic emission from the nanocrystal film (green circle), and then calculating the solid angle that enters the eyepiece lens. The power is scaled by the transmittance of each optical element through which the light passes.

**Table S1:** $\eta_{expected}$ for the down-conversion imaging setup with compounded improvements from the measured conditions described in the main text.

| | Measured | Lossless optics (no reflections) | Back reflector | EQE = 20% Abs = 7.5% → 100% | EQE = 100% | NA = 0.8 |
|---|---|---|---|---|---|---|
| $\eta$ (%) | 0.00133 | 0.00203 | 0.0046 | 0.159 | 0.797 | 40.1 |

## 4. FDTD far field of nanocrystal photoluminescence

We performed 3D full-wave simulations using Lumerical FDTD to determine the percent of power that can outcouple from the nanocrystal film and substrate into free space such that it can be captured by the objective optics. Figure S5 shows the cross section of the simulation. Perfectly-matched-layer (PML) absorbing boundary conditions were used to define the simulation area (figure S4a). A 510-nm-wavelength emitting dipole was placed in the middle of an 80-nm-thick layer of material with a refractive index, n, set to 2.1—the same value as CsPbBr$_3$ film at a wavelength of 510 nm. A transmission field monitor was placed in the glass substrate a few hundred nanometers above the CsPbBr$_3$-glass interface to record the near-field profile. The far-field intensity profile was calculated in the glass for the three dipole polarization orientations (figure S4b-d) using a far-field transform. The power that escapes the glass was calculated by integrating the far-field intensity over the critical angle for the glass-air interface ($\theta_c = 40.5°$) and scaled by the s- and p-polarized transmittances for all incident angles between the glass-air interface. The ratio between the escape power and the dipole source power quantifies the amount of converted light that can radiate towards the objective optics, equivalent to the outcoupling efficiency $QY_{out}$. The x- and y-polarized simulations had a ratio of 6.4% each. The z-polarized simulation had a ratio of 3%. The average of all three polarizations gives the unpolarized ratio of to be 5.3% of the source power escaping into free space.

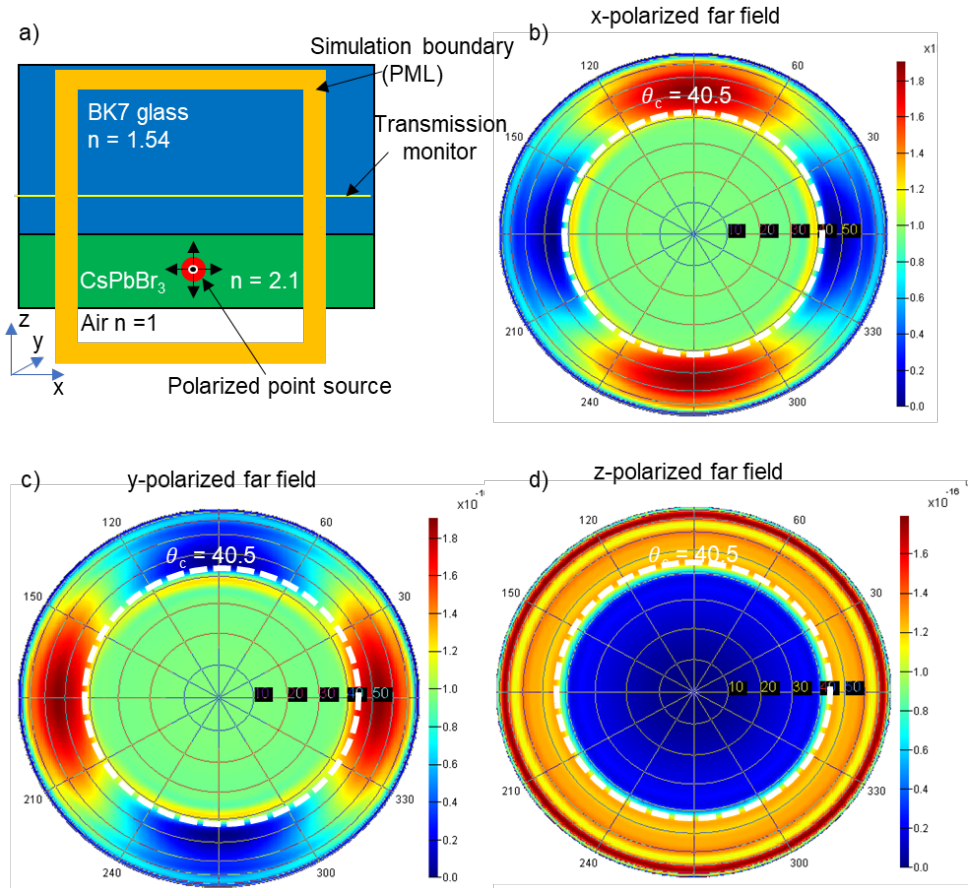

**Figure S5: Full-wave simulations.** a) Cross-sectional diagram of the 3D simulation space replicating the conditions of our nanocrystal-on-glass sample. A point source emitting 510 nm is used to represent an emitting nanocrystal and is placed in the middle of the 80-nm-thick layer representing the CsPbBr$_3$ nanocrystal film with n = 2.1. The source is polarized along 3 axial directions. A transmission field monitor is used to record the electric fields that enter the BK7 glass substrate. b-d) The far-field profiles of the electric field intensity of light radiating into the glass substrate for each axial polarization of the source. Highlighted in dashed white is the critical angle, $\theta_c$ for which light can escape into free space at the glass-air interface towards the direction of the eyepiece lens.